\documentclass[twocolumn,pra,showpacs,superscriptaddress]{revtex4-1}
\usepackage{graphicx}
\usepackage{subfigure}
\usepackage{amsmath}
\usepackage{amsfonts,amssymb}
\usepackage{bm}
\usepackage{float}
\usepackage{color}
\usepackage{sidecap}
\allowdisplaybreaks
\usepackage{soul}
\setstcolor{red}
\usepackage{ulem}
\usepackage{hyperref}
\hypersetup{pdfpagemode=FullScreen,colorlinks=true,breaklinks,urlcolor=blue,linkcolor=blue,citecolor=blue}

\begin{document}

\title{Generalized Aubry-Andr\' e-Harper model with p-wave superconducting pairing}

\author{Qi-Bo Zeng}
\email{zqb15@mails.tsinghua.edu.cn}
\affiliation{Department of Physics and State Key Laboratory of Low-Dimensional Quantum Physics, Tsinghua University, Beijing 100084, China}

\author{Shu Chen}
\affiliation{Beijing National Laboratory for Condensed Matter Physics, Institute of Physics,
Chinese Academy of Sciences, Beijing 100190, China}
\affiliation{Collaborative Innovation Center of Quantum Matter, Beijing, China}

\author{Rong L\"u}
\email{rlu@tsinghua.edu.cn}
\affiliation{Department of Physics and State Key Laboratory of Low-Dimensional Quantum Physics, Tsinghua University, Beijing 100084, China}
\affiliation{Collaborative Innovation Center of Quantum Matter, Beijing, China}

\begin{abstract}
We investigate a generalized Aubry-Andr\' e-Harper (AAH) model with p-wave superconducting pairing. Both the hopping amplitudes between the nearest neighboring lattice sites and the on-site potentials in this system are modulated by a cosine function with a periodicity of $1/\alpha$. In the incommensurate case [$\alpha=(\sqrt{5}-1)/2$], due to the modulations on the hopping amplitudes, the critical region of this quasiperiodic system is significantly reduced and the system becomes more easily to be turned from extended states to localized states. In the commensurate case ($\alpha = 1/2$), we find that this model shows three different phases when we tune the system parameters: Su-Schrieffer-Heeger (SSH)-like trivial, SSH-like topological, and Kitaev-like topological phases. The phase diagrams and the topological quantum numbers for these phases are presented in this work. This generalized AAH model combined with superconducting pairing provides us with a useful testfield for studying the phase transitions from extended states to Anderson localized states and the transitions between different topological phases.
\end{abstract}


\maketitle
\date{today}

\section{Introduction}

The Aubry-Andr\' e-Harper (AAH) model has been extensively used as a quasiperiodic model to theoretically study the phase transition between extended, critical and localized phases \cite{Harper, Aubry, Hiramoto1, Ostlund, Kohmoto1, Kohmoto2, Thouless, Hiramoto2, Geisel, Han, Chang, Takada, Liu, Wang}. With the realization of the AAH model in photonic crystals \cite{Negro, Lahini, Kraus1} and ultracold atoms \cite{Roati, Modugno}, this model has gained attention in recent years. The abundant phenomena revealed by the AAH model make it an ideal test field for topological phases and the transitions between them both in the incommensurate and commensurate cases \cite{Kraus2, Lang1, Lang2, Cai, DeGottardi, Ganeshan, Satija, Barnett, Deng, Grusdt, Zhu}. When the lattice is incommensurate, the system will go through a transition from an extended phase to a localized phase due to the disordered on-site potential \cite{Cai, DeGottardi}. If the lattice is commensurate, the AAH model can be used to explore emerging topological states of matter. In addition, quantum many-body localization phenomena are also widely explored in this model \cite{Iyer, Schreiber}. Recently, the connection between the AAH model, the one-dimensional Kitaev chain \cite{Kitaev}, and topological superconductivity are investigated. In Ref. \cite{Ganeshan}, a generalized AAH model which has modulations both in the on-site potentials and the hopping amplitudes between the neighboring lattice sites is discussed, and the topological states are attributed to the topological properties of the one-dimensional Majorana chain. This model is also connected to the Su-Schrieffer-Heeger (SSH) \cite{Su} model when the lattice is commensurate. In Ref. \cite{Wang}, the phase diagram of a non-Abelian AAH model with p-wave superfluidity is discussed and the phase transition from a metallic phase to an insulator phase is checked. However, most AAH models explored before have only modulations in the on-site potential or do not include the superconducting pairing, thus limiting the usage of these models. If we introduce superconducting pairing into the generalized AAH models discussed in Ref. \cite{Ganeshan}, more interesting phenomena can be expected. With p-wave superconducting pairing and modulations in the model, there would be both SSH-like and Kitaev-like topologically nontrivial phases, and thus so-called electron fractionalization to the Majorana fermions can be realized, as discussed in Ref. \cite{Wakatsuki}.

In this paper, we consider a generalized AAH model with a superconducting pairing potential. Both the hopping amplitudes between the nearest neighboring lattice sites and the on-site potentials are modulated by a cosine function which has a periodicity of $1/\alpha$. When $\alpha$ is irrational, the lattice is incommensurate, and we find that the system will go through a phase transition from the extended states to localized states when the disordered on-site potential increases beyond some critical value. This critical value as well as the critical region of the system will be significantly reduced due to the modulation in the hopping amplitude and thus makes the system much easier to be localized. On the other hand, in the commensurate lattice when $\alpha$ is rational, this model shows different topological phases as we tune the system parameters. We mainly discuss the case with $\alpha=1/2$ and find that there are three phases presented in the system: a Su-Schrieffer-Heeger (SSH)-like trivial phase, a SSH-like topological phase, and a Kitaev-like topological phase. By calculating the topological numbers of the system, we present the phase diagrams and phase boundaries for these phases. This generalized AAH model provides us with a useful test field for the study of phase transitions between extended states and Anderson localized states, or the transitions between different topological phases.

The rest of the paper is organized as follows. In Sec. \ref{sec2}, we introduce the generalized AAH model and write down the Hamiltonian. Then we consider the incommensurate case in Sec. \ref{sec3}, where we will discuss the phase transition from the extended phase to the localized phase, and especially the influences of the modulations in the hopping amplitude on this transition. In Sec. \ref{sec4}, we check the commensurate lattice with $\alpha=1/2$ and investigate those different phases presented in this system. The last section (Sec. \ref{sec5}) is dedicated to a brief summary.

\section{Model Hamiltonian}\label{sec2}
The Hamiltonian of the generalized one-dimensional (1D) Aubry-Andr\' e-Harper model we consider in this paper is described by the following Hamiltonian

\begin{equation}\label{Eq1}
H= \sum_{j=1}^{N} V_j c_j^\dagger c_j + \sum_{j=1}^{N-1} [ -t_j c_{j+1}^\dagger c_j +\Delta c_{j+1}^\dagger c_j^\dagger + H.c. ]
\end{equation}
where $c_j^\dagger$ ($c_j$) is the creation (annihilation) operator at site $j$, $V_{j} = V \cos(2\pi \alpha j + \varphi_V)$ is the on-site potential, $t_j = t [1+\lambda \cos (2\pi \alpha j + \varphi _\lambda)]$ is the hopping amplitude between the nearest neighboring lattice sites, and $\Delta$ is the superconducting pairing gap which is taken to be real. This one-dimensional chain has $N$ sites and both the on-site potential and the hopping amplitude are modulated by a cosine function with periodicity $1/\alpha$ but with phase factor $\varphi_v$ and $\varphi_\lambda$ , respectively. If $\Delta=0$, this model will reduce to the generalized AAH model introduced in Ref. \cite{Ganeshan}, which covers both the diagonal and off-diagonal AAH model. If $\lambda=0$, then the Hamiltonian is the same as the 1D Kitaev chain with on-site potential modulations which are investigated in detail in Refs. \cite{Cai, DeGottardi}. It has been predicted that if $\alpha$ is irrational, the Kitaev model $(\lambda=0, V \neq 0)$ or the diagonal AAH model $(\lambda=0, \Delta=0, V \neq 0)$ will go through a localization transition where all extended states become localized as $V$ is increased beyond some critical value. However, if $\alpha$ is rational, it is known that there is no such phase transition in these two models. For simplicity, we will set $\varphi_V = \varphi_\lambda = \varphi$ in this paper but we emphasize that all the results obtained in this paper can also be extended to the $\varphi_V \neq \varphi_\lambda$ case.

By introducing Majorana operators $c_j = (\gamma_{j,1} + i \gamma_{j,2})/2$, the Hamiltonian could be rewritten as

\begin{equation}\label{Eq2}
\begin{split}
H &= \frac{i}{2} \sum_{j} \{ V_{2j-1} \gamma_{2j-1,1} \gamma_{2j-1,2} + V_{2j} \gamma_{2j,1} \gamma_{2j,2} \\
&+ ( -t_{2j-1} - \Delta ) \gamma_{2j,1} \gamma_{2j-1,2} + ( -t_{2j} + \Delta ) \gamma_{2j,1} \gamma_{2j+1,2} \\
&+ ( t_{2j-1} - \Delta ) \gamma_{2j,2} \gamma_{2j-1,1} + ( t_{2j} + \Delta ) \gamma_{2j,2} \gamma_{2j+1,1} \}.
\end{split}
\end{equation}
Thus the system can be taken as two 1D Majorana chains coupled by potential $V_j$ and we may expect that there will be different topological phases depending on whether or not the Majorana fermions at the ends of the Majorana chains are paired. The phase transition from the extended state to the localized state could also be expected according to previous works on the quasiperiodic 1D Kitaev chain \cite{Cai, DeGottardi}. In the following, we will discuss the properties of this generalized AAH model both in the incommensurate case and the commensurate case.

\section{Incommensurate case}\label{sec3}

When $\alpha$ is irrational, the model system becomes quasiperiodic. Since the model is similar to a 1D Kitaev model with modulated on-site potentials which shows an Anderson localization transition when $V$ becomes large enough [$V > 2(t+\Delta)$], we may expect that the same transition would also show up in our system. We take $\alpha = (\sqrt{5}-1)/2$ as an example, but the results can also be generalized to other incommensurate situations. The Hamiltonian can be diagonalized by using the Bogoliubov-de Gennes (BdG) transformation \cite{Gnnes, Lieb},
\begin{equation}\label{}
  \eta_n^\dagger = \sum_{j=1}^{N} [u_{n,j} c_{n,j}^\dagger + v_{n,j} c_{n,j}],
\end{equation}
where $u_{n,j}$ and $v_{n,j}$ are chosen to be real and $n$ is the energy band index. Then the wave function of the Hamiltonian is
\begin{equation}\label{}
  | \Psi_n \rangle = \eta_n^\dagger | 0 \rangle =  \sum_{j=1}^{N} [u_{n,j} c_{n,j}^\dagger + v_{n,j} c_{n,j}] | 0 \rangle.
\end{equation}
From the Schr\"odinger equation $\mathcal{H} | \Psi_n \rangle = \epsilon_n | \Psi_n \rangle$, we have
\begin{widetext}
\begin{equation}
\left\{
  \begin{aligned}
  -t_{j-1} u_{n,j-1} + \Delta v_{n,j-1} + V_j u_{n,j} - t_j u_{n,j+1} - \Delta v_{n,j+1} &= \epsilon_n u_{n,j},\\
  -\Delta u_{n,j-1} + t_{j-1} v_{n,j-1} - V_j v_{n,j} + \Delta u_{n,j+1} + t_j v_{n,j+1} &= \epsilon_n v_{n,j}.
  \end{aligned}
 \right.
\end{equation}
By representing the wave function as
\begin{equation}\label{}
  | \Psi_n \rangle = [u_{n,1}, v_{n,1}, u_{n,2}, v_{n,2}, \cdots, u_{n,N}, v_{n,N}]^T,
\end{equation}
the Hamiltonian can be written as a $2N \times 2N$ matrix,
\begin{equation}\label{}
\mathcal{H}_n=
  \begin{pmatrix}
    A_1 & B & 0 & \cdots & \cdots & \cdots & C \\
    B^\dagger & A_2 & B & 0 & \cdots & \cdots & 0 \\
    0 & B^\dagger & A_3 & B & 0 & \cdots & 0 \\
    \vdots & \ddots & \ddots & \ddots & \ddots & \ddots & \vdots \\
    0 & \cdots & 0 & B^\dagger & A_{N-2} & B & 0 \\
    0 & \cdots & \cdots & 0 & B^\dagger & A_{N-1} & B \\
    C^\dagger & \cdots & \cdots & \cdots & 0 & B^\dagger & A_N
  \end{pmatrix},
\end{equation}
\end{widetext}
where
\begin{equation}\label{}
  A_j =
  \begin{pmatrix}
    V_j & 0 \\
    0 & -V_j
  \end{pmatrix},
\end{equation}
\begin{equation}\label{}
  B=\begin{pmatrix}
      -t_j & -\Delta \\
      \Delta & t_j
    \end{pmatrix},
\end{equation}
and
\begin{equation}\label{}
  C=\begin{pmatrix}
      -t_{j+1} & \Delta \\
      -\Delta & t_{j+1}.
    \end{pmatrix}
\end{equation}
Here we have assumed a periodic boundary condition which implies that $c_{j+N}=c_j$. This is legitimate when $N$ is large enough since the irrational number $\alpha$ can be approximated by rational numbers \cite{Wang}. From this Hamiltonian, we can get $u_{n,j}$ and $v_{n,j}$ and thus the wave function of the system.

In order to investigate the phase transition of this incommensurate 1D chain, we need to calculate the inverse participation ratio (IPR) which for a normalized wave function $\Psi_n$ is defined as $IPR=\sum_j (u_{n,j}^4 + v_{n,j}^4)$. The IPR measures the inverse of the number of occupied lattice sites and is a very useful quantity in characterizing the localization transitions of quasiperiodic systems. We further define the mean inverse participation ratio (MIPR) as $MIPR = \frac{1}{2N} \sum_{n=1}^{2N} \sum_{j=1}^{N}(u_{n,j}^4 + v_{n,j}^4)$. MIPR is close to zero for extended states of the system, however, it will tend to a finite value of $O(1)$ for localized states.

\begin{figure}[!ht]
\centering
\subfigure[MIPR vs V]{
\label{fig1a}
\includegraphics[width=3.1in]{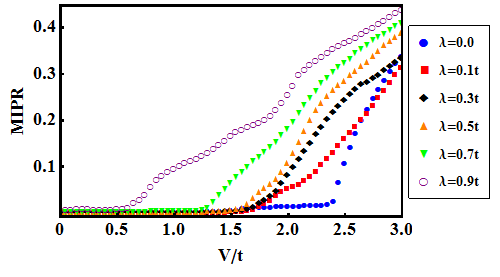}}
\subfigure[MIPR vs $\lambda$]{
\label{fig1b}
\includegraphics[width=3.1in]{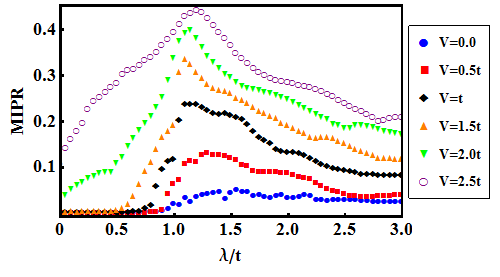}}
\caption{(Color online) (a) MIPR vs V under different $\lambda$; (b) MIPR vs $\lambda$ under different V. Here we set $\Delta=0.2t$, $\varphi=0$, and the number of lattice sites $N=377$, which is the 14th Fibonacci number.}
\label{fig1}
\end{figure}

Figure \ref{fig1} shows the variation of the MIPR as the parameter of the system changes. Throughout this paper, we take $t=1$ as the energy unit. Other parameters in Fig. \ref{fig1} are $\Delta=0.2t$, $\varphi=0$, and the number of lattice sites $N=377$. The solid blue dots in Fig.\ref{fig1a} corresponds to the usual 1D Kitaev chain with a quasiperiodic disordered on-site potential ($\lambda=0, V \neq 0$). The phase transition takes place when $V_c>2(t+\Delta)$, as was reported in earlier work \cite{Cai}. There is a plateau in the MIPR before it increases sharply when $\lambda=0$, which corresponds to the critical phase before the system enters the localized phase, which is discussed in Ref. \cite{Wang}. If $\lambda$ becomes nonzero and gets stronger, as is shown in Fig. \ref{fig1a}, the critical value of the phase transition $V_c$ is reduced and the critical region will disappear gradually. It becomes more significant when $\lambda$ is close to the value of $t$, as shown by the triangular green dots and the open purple dots in Fig. \ref{fig1a}, where the MIPR increases rapidly at a much smaller value of $V$. This is understandable since as $\lambda$ increases, especially when it reaches the strength of $t$, the modulation in the hopping amplitude will become increasingly stronger, the coupling strength between some lattice sites then will be weakened and the system thus can be localized more easily and more quickly. In Fig. \ref{fig1b}, we present the evolution of the MIPR as a function of $\lambda$ with different $V$. From this figure, we can see that when $V=0$, the MIPR will increase from zero to a finite though small value, which means that even without the disorders coming from the on-site potentials, the modulation in the hopping amplitude alone can also influence the extended state of the system. When $V \neq 0$, it is clear that the MIPR increases sharply when $\lambda$ is close to $1$ and reaches to a maximum, then drops again when $\lambda$ becomes stronger. This signifies that the system is more likely to be localized when the hopping amplitudes are also modulated, which is consistent with the results we get in Fig. \ref{fig1a}.

So due to the modulation in the hopping amplitude between the nearest neighboring lattice sites, we find that the incommensurate generalized AAH model with p-wave superconducting pairing is easier to be localized when tuning the disordered on-site potential. The critical value of the potential beyond which the phase transition happens is significantly reduced especially when the modulation in the hopping amplitude is strong. These results can also be generalized to situations with other irrational $\alpha$ values.

\section{Commensurate case}\label{sec4}

When $\alpha$ is rational, the lattice is commensurate. It is well known that the system will not undergo a localization transition as that in the incommensurate case. In this section, we mainly discuss the $\alpha=1/2$ case. The generalized AAH model with p-wave superconducting pairing in this case can be tuned between different topological phases including the Su-Schrieffer-Heeger-like (SSH-like) topological phase and the Kitaev-like topological phase. When $\alpha=1/2$, the Hamiltonian in Eq. (\ref{Eq1}) becomes
\begin{widetext}
\begin{equation}\label{}
  \begin{split}
     H= &-V \cos \varphi \sum_j ( c_{2j-1}^\dagger c_{2j-1} - c_{2j}^\dagger c_{2j} ) -t \sum_j [(1-\lambda \cos \varphi) c_{2j}^\dagger c_{2j-1} + (1+\lambda \cos \varphi) c_{2j+1}^\dagger c_{2j} + H.c. ] \\
       & +\Delta \sum_j ( c_{2j}^\dagger c_{2j-1}^\dagger + c_{2j+1}^\dagger c_{2j}^\dagger + H.c. ) \\
       =& -V \cos \varphi \sum_j ( c_{A,j}^\dagger c_{A,j} - c_{B,j}^\dagger c_{B,j} ) - t \sum_j [ (1-\lambda \cos \varphi) c_{B,j}^\dagger c_{A,j} + (1+\lambda \cos \varphi ) c_{A,j+1}^\dagger c_{B,j} + H.c. ]\\
       &+\Delta \sum_j ( c_{B,j}^\dagger c_{A,j}^\dagger + c_{A,j+1}^\dagger c_{B,j}^\dagger + H.c. ).
  \end{split}
\end{equation}
\end{widetext}
This Hamiltonian is very similar to the one discussed in Ref. \cite{Wakatsuki}, except that here the modulations are added to the on-site potential and the hopping amplitude by the function $\lambda \cos \varphi$. It is reduced to the SSH model when $\mu=\Delta=0$ and to the 1D Kitaev model when $\lambda=0$ except that the chemical potential is positive or negative alternatively. Since both SSH-like and Kitaev-like models are included in this generalized system, we may expect that with appropriate system parameters, different topological phases would show up. To verify this we can check the energy spectrum of the system, which can be reached by diagonalizing the Hamiltonian rewritten in the Majorana fermions representation [see Eq. (\ref{Eq2}))].

\begin{figure}[!ht]
\centering
\includegraphics[width=3.1in]{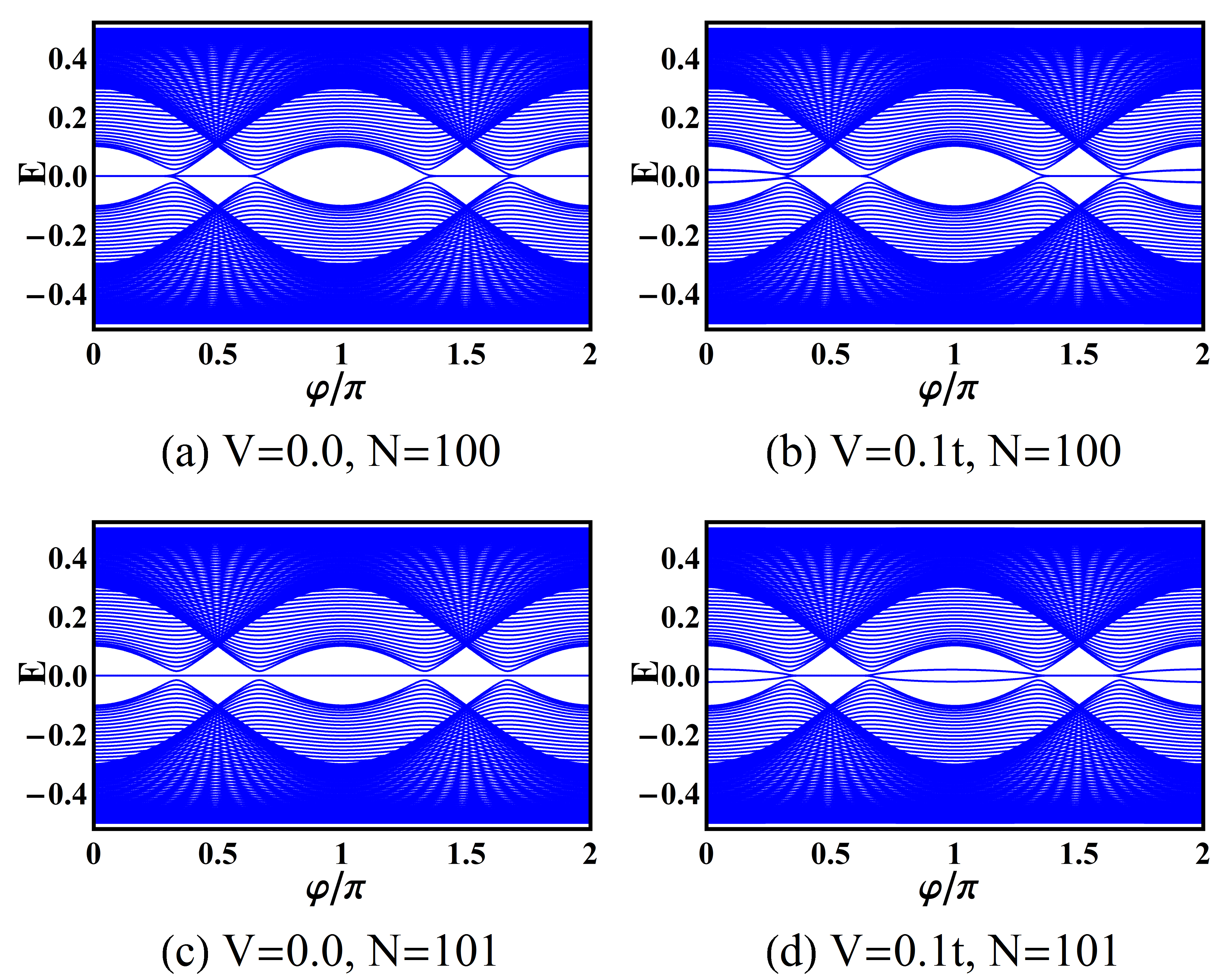}
\caption{(Color online) Energy vs $\varphi$ with different on-site chemical potential $V$ and different site numbers. In (a) and (b), the site number is chosen to be even, $N=100$, while in (c) and (d), it is chosen to be odd, $N=101$. Other parameters are $\lambda=0.4t$, $\Delta=0.2t$. The zero energy modes in the gap are doubly degenerate when $V$ is zero. However, if $V$ takes nonzero values, some of the zero modes split and become nondegenerate. The regions with zero energy modes unchanged when $V$ is nonzero correspond to the Kitaev-like topological phase, while those regions with zero energy modes splitting in the finite $V$ situation correspond to the SSH-like topological phase.}
\label{fig2}
\end{figure}

\begin{figure}[!ht]
\centering
\includegraphics[width=3.1in]{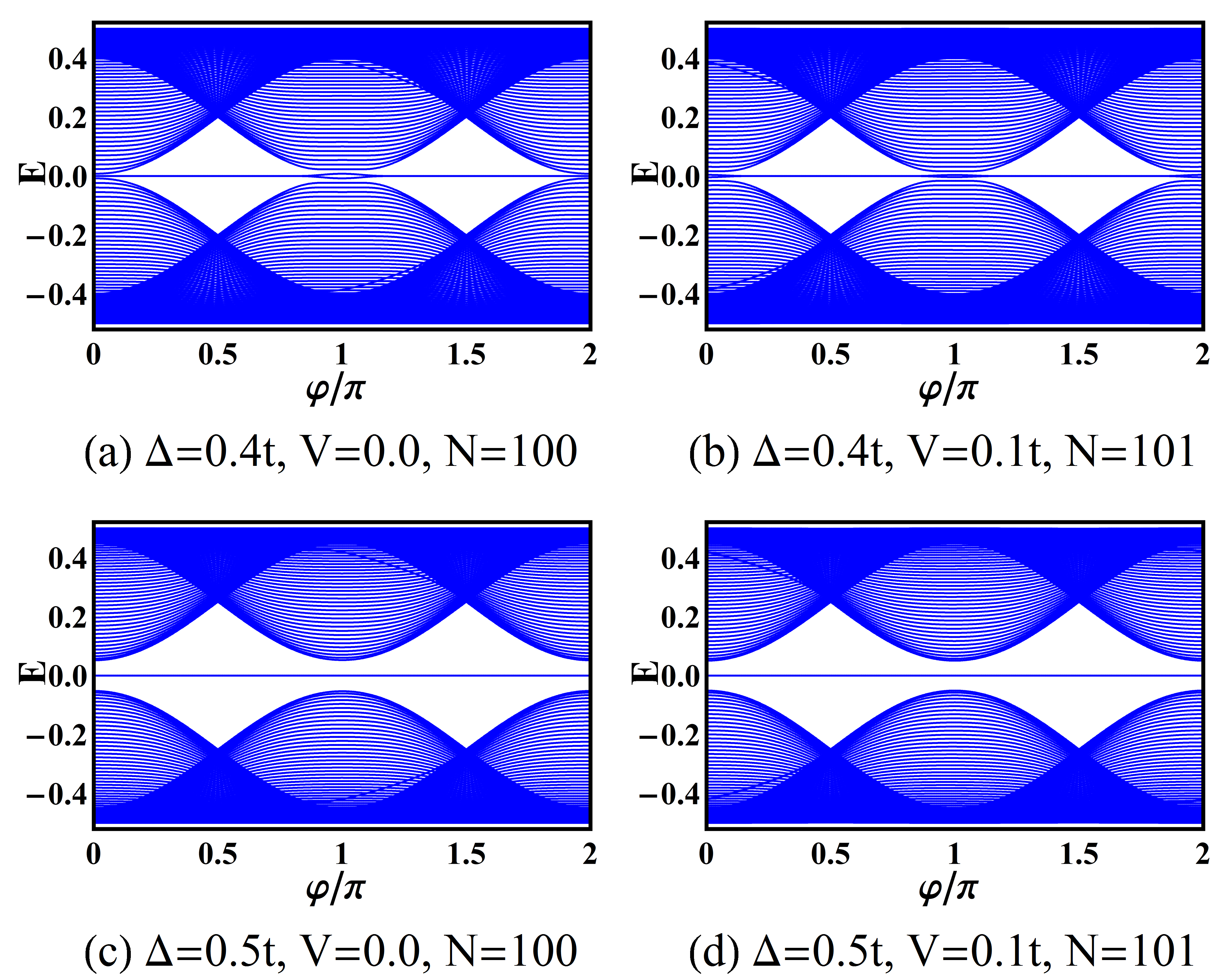}
\caption{(Color online) Energy vs $\varphi$ with different on-site chemical potential $V$ and different site numbers. $\lambda$ is set to be $0.4t$. (a) $\Delta=0.4t$, $V=0.0$, $N=100$; (b) $\Delta=0.4t$, $V=0.1t$, $N=101$; (c) $\Delta=0.5t$, $V=0.0$, $N=100$; and (d) $\Delta=0.5t$, $V=0.1t$, $N=101$.}
\label{fig3}
\end{figure}

In Fig. \ref{fig2}, we present the spectra of the system with an even site number ($N=100$) and an odd site number ($N=101$). From Fig. \ref{fig2}(a), we can see that there are gapped topologically trivial regimes and nontrivial regimes with zero energy modes. However, comparing with Fig. \ref{fig2}(b), it is clear that when $V$ becomes nonzero, parts of the zero modes in the topologically nontrivial regime split into two non zero energy modes, but the rest of the zero modes remain unchanged. More interestingly, when the site number is odd, there are also zero modes in the trivial regime when the site number is even and those zero modes also split when $V$ becomes nonzero [see Fig. \ref{fig2}(c) and \ref{fig2}(d)]. This even-odd effect in the energy spectrum is a feature of the SSH model. The regime with zero energy modes which will not split when $V$ is finite is the Kitaev topological regime. So there are three phases in this generalized system: a SSH-like trivial phase, a SSH-like topological phase, and a Kitaev-like topological phase. In the SSH-like topological phase, there will be one Dirac fermion at each end of the chain, while in the Kitaev-like topological phase, there will be one Majorana fermion at each end of the chain. The zero energy modes originate from the particle-hole symmetry which has two sources: In the SSH-like phase, it is the sublattice symmetry (when $V=0$) while in the Kitaev-like phase, it is the superconductivity. As the superconducting pairing strength becomes stronger, the superconductivity will dominate and there will be no SSH-like phase in the system. Figure \ref{fig3} shows the energy spectra of the system with $\Delta =0.4t$ and $0.5t$ both in the even and odd site number situations. When $\Delta=0.4t$, there is only a very small region which shows the even-odd effect, as shown in Figs. \ref{fig3}(a) and \ref{fig3}(b). When $\Delta$ becomes larger, there are always zero modes as long as the system is in the topological nontrivial phase [Fig. \ref{fig3}(c) and \ref{fig3}(d)]. So the Kitaev-like phase dominates and the SSH-like phase disappears. As the parameters change, there would be a phenomenon of electron fractionization to the Majorana fermions \cite{Wakatsuki}.

\begin{figure}[!ht]
\centering
\includegraphics[width=3.2in]{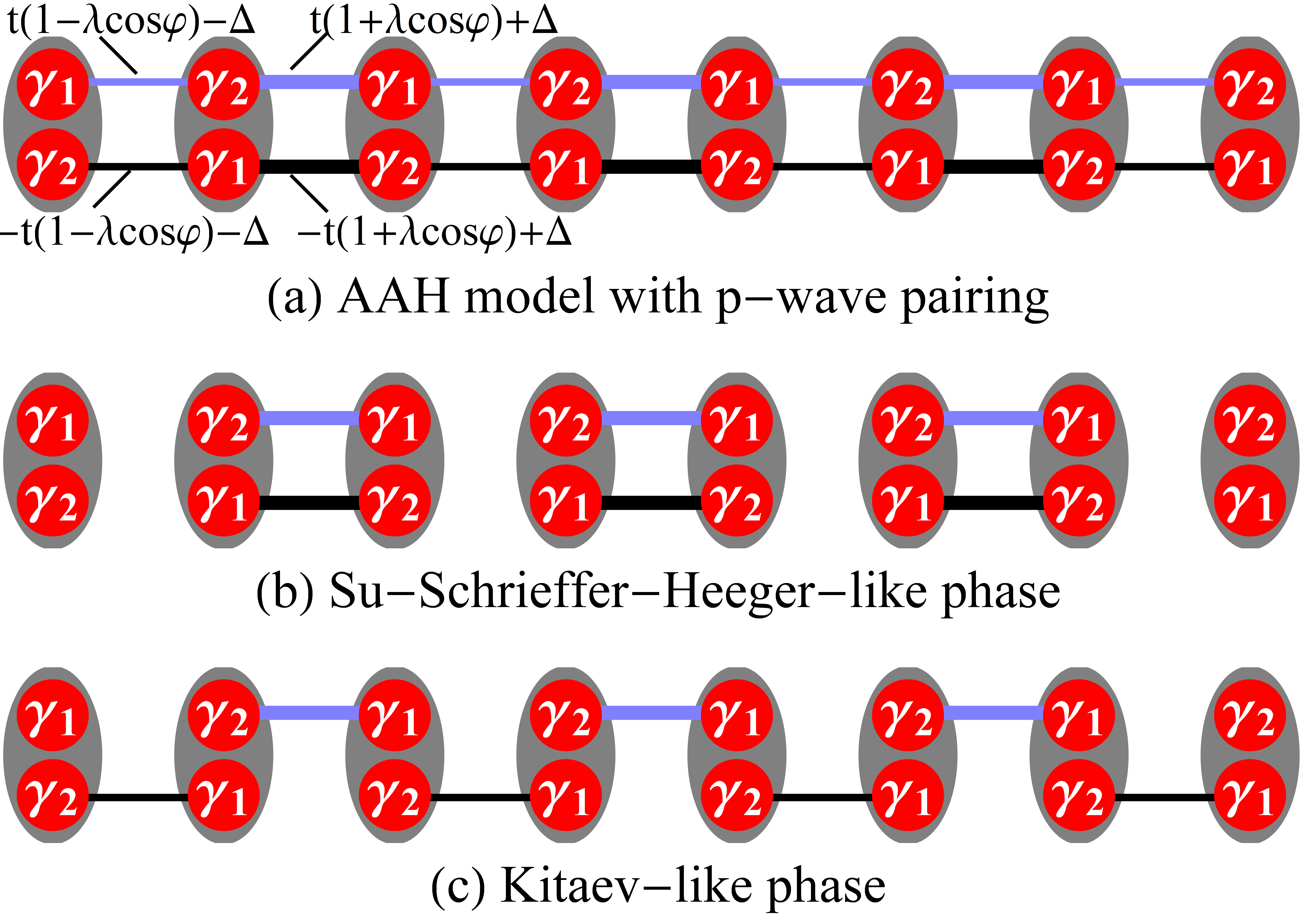}
\caption{(Color online) Illustrations of the AAH model and the different topological phases in the commensurate case. (a) The AAH model can be taken as two coupling Majorana fermion chains according to Eq. (\ref{Eq2}) (here we set $\alpha=1/2$). $\gamma_1$ and $\gamma
_2$ are the Majorana fermions composing the fermionic particle at each lattice site. The bonds with different colors and thicknesses represent the coupling strength between the Majorana fermions. (b) SSH-like topological phase with an unpaired fermion at each end of the chain. (c) In the Kitaev-like topological phase, there is an unpaired Majorana fermion at each end of the chain.}
\label{fig4}
\end{figure}

To give a more intuitive picture of the different topological phases in this commensurate situation, we illustrate the schematic diagrams of the AAH model in Fig. \ref{fig4} by setting $\alpha=1/2$ in Eq. (\ref{Eq2}). As shown by Fig. \ref{fig4}(a), the AAH model can be taken as two Majorana fermion chains which are coupled by the on-site chemical potentials. The different bonding strengths between the Majorana fermions will lead to different topological phases. When the system is in the SSH-like topological phase, there is a fermion at each end of the one-dimensional chain, [see Fig. \ref{fig4}(b)]. If the system is in the Kitaev-like topological phase, there will be an unpaired Majorana fermion at each end of the chain, which is shown in Fig. \ref{fig4}(c).

In order to characterize these different topological phases and determine the phase boundaries, we need to calculate the topological numbers. First, we transform the Hamiltonian into the momentum representation. By defining
\begin{equation*}
  \Psi = [c_{qA}^\dagger  \ c_{qB}^\dagger \ c_{-qA} \ c_{-qB}]^T,
\end{equation*}
the Hamiltonian can be written as $\mathcal{H}=\frac{1}{2} \sum_q \Psi^\dagger \mathcal{H} (q) \Psi$, where
\begin{equation}\label{}
  \mathcal{H} (q)=
  \begin{pmatrix}
    -V\cos \varphi & g(q) & 0 & h(q) \\
    g^*(q) & V\cos \varphi & -h^*(q) & 0 \\
    0 & -h(q) & V\cos \varphi & -g(q) \\
    h^*(q) & 0 & -g^*(q) & -V\cos \varphi
  \end{pmatrix}
\end{equation}
with $g(q) = -t [ (1-\lambda \cos \varphi) + (1+ \lambda \cos \varphi) e^{-iq} ]$ and $h(q) = -\Delta (1 - e^{-iq})$. After diagonalizing $\mathcal{H} (q)$ the eigenvalues can be obtained as
\begin{equation}\label{}
 \begin{split}
  E^2&(q) = (V \cos \varphi)^2 + |g(q)|^2 + |h(q)|^2 \\
  & \pm 2\sqrt{(V\cos \varphi)^2 |h(q)|^2 + 4\Delta^2 t^2 (\lambda \cos \varphi)^2 (\cos q -1)^2}.
  \end{split}
\end{equation}
Thus we have
\begin{equation}\label{eigenvalues}
  \begin{split}
    E(0)=&\pm \sqrt{(V\cos \varphi)^2 + 4t^2}, \\
    E(\frac{\pi}{2}) =& ( \sqrt{(V\cos \varphi)^2 + 2t^2 (\lambda \cos \varphi)^2} \pm \sqrt{2} \Delta)^2 + 2t^2, \\
    E(\pi) =& \pm \sqrt{(V\cos \varphi)^2 + 4t^2 (\lambda \cos \varphi)^2} \pm 2\Delta.
  \end{split}
\end{equation}
The gap will close at $q=\pi$ when $(V\cos \varphi)^2 = 4[\Delta^2 - t^2 (\lambda \cos \varphi)^2]$. For $\Delta=0$, the spectrum is reduced to
\begin{equation}\label{}
  \begin{split}
  &E(q)=\\
  &\pm \sqrt{V^2 \cos^2 \varphi + 2t^2 [ 1+\lambda^2 \cos^2 \varphi + (1-\lambda^2 \cos^2 \varphi) \cos q ]}
  \end{split}
\end{equation}
which is similar to the SSH model spectrum.  If $\lambda=0$, then the spectrum is reduced to
\begin{equation}\label{}
  E(q) = \pm \sqrt{(V\cos \varphi \pm 2\Delta \sin \frac{q}{2})^2 + 4t^2 \cos^2 \frac{q}{2}},
\end{equation}
which is similar to the Kitaev model spectrum. The difference here is that the system is topological for $|V \cos \varphi| <2|\Delta|$ and trivial for $|V \cos \varphi|>2|\Delta|$. However, the spectrum for a usual Kitaev model is $E(q)=\pm \sqrt{(2t\cos \frac{q}{2} - \mu)^2 + 4 \Delta^2 \sin^2 \frac{q}{2}}$, where the system is topological for $|\mu|<2t$ and trivial for $|\mu|>2t$. In order to make further investigations about the topological phases of the commensurate lattice, we need to check the topological quantum numbers of the system with different parameters.

\begin{figure}[!ht]
\centering
\includegraphics[width=3.1in]{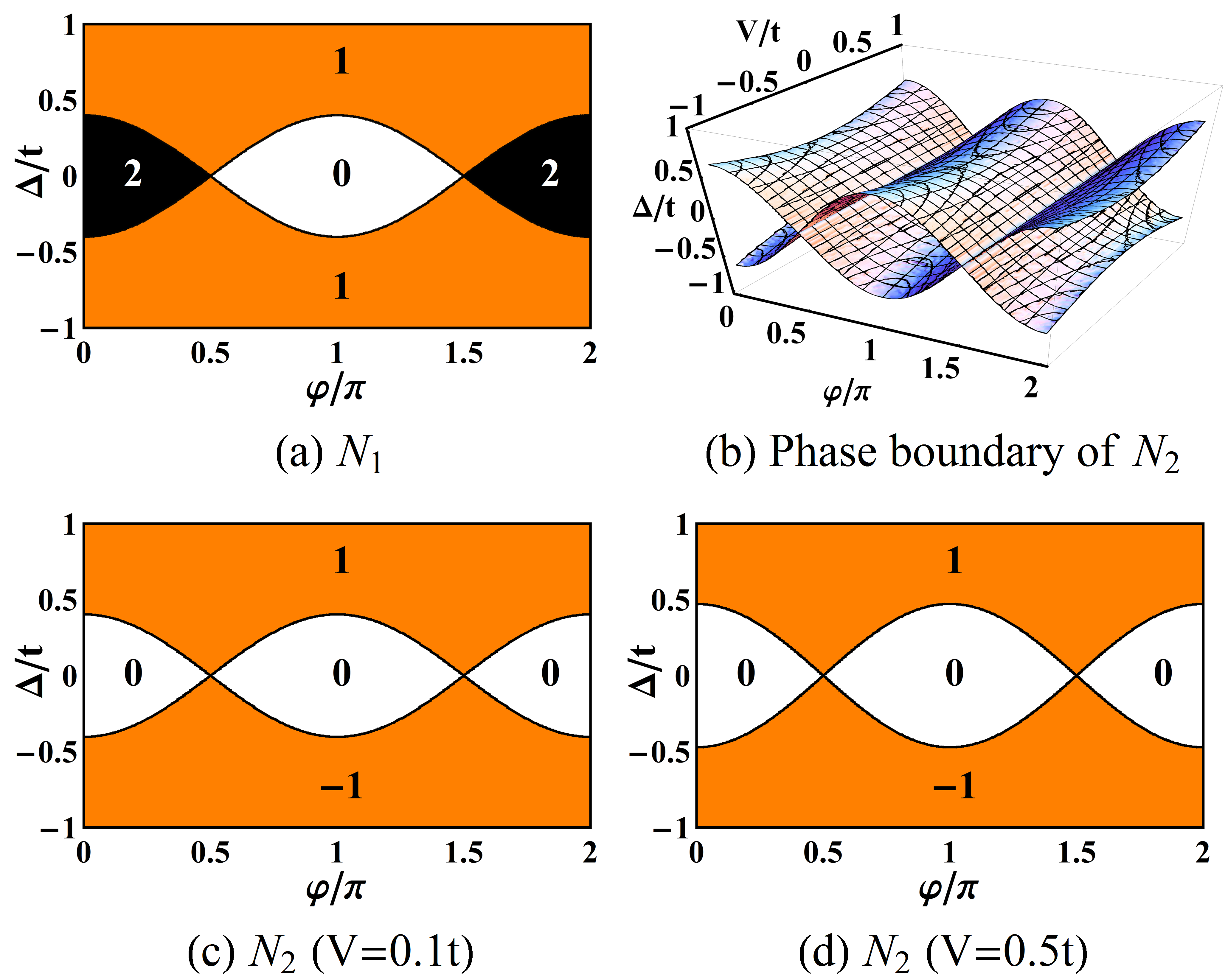}
\caption{(Color online) (a) Phase diagram of the system in the $V=0$ case; (b) phase boundary of the system in the $V\neq 0$ case; (c) and (d) show the cross sections of (b) at $V=0.1t$ and $V=0.5t$, respectively. The numbers in the figures denote the topological number $N_1$ or $N_2$. Here $\lambda$ is also set to be $0.4t$.}
\label{fig5}
\end{figure}

Following the discussions in Ref. \cite{Wakatsuki}, we find that the topological class of our model is also BDI and we can characterize our system by the $\mathbb{Z}$ index. Now we calculate the topological numbers of the system. Firstly, we consider the sublattice symmetric case ($V=0$) whose topological number can be calculated as (check the Appendix for details)
\begin{equation}\label{}
  N_1 = \Theta (t \lambda \cos \varphi -\Delta) + \Theta (t\lambda \cos \varphi +\Delta)
\end{equation}
with $\Theta(x)$ being the step function. Figure \ref{fig5}(a) illustrates the phase diagram of the system according the value of topological number $N_1$. There are three phases: (i) When $|\Delta|< t \lambda |\cos \varphi|$, $\cos \varphi < 0$, $N_1=0$, the system is in the SSH-like trivial phase; (ii) when $|\Delta| < t\lambda |\cos \varphi|$, $\cos \varphi >0$, $N_1 =2$, the system is in the SSH-like topological phase; (iii) when $|\Delta|>t\lambda |\cos \varphi|$, $N_1=1$, the system is in the Kitaev-like topological phase. As shown in this figure, when $|\Delta| > \lambda$ (here $\lambda=0.4t$), the system will be totally in the Kitaev-like topological phase, which is consistent with the numerical results for the energy spectra we presented in Fig. \ref{fig3}. The topological number $N_1$ can be interpreted as the number of Majorana fermion pairs at the ends of the chain. So when $N_1=2$, there will be two Majorana fermions at each end of the chain which will pair up with each other and become a Dirac fermion, which is the feature of the SSH model. If $N_1=1$, there will be one single unpaired Majorana fermion at each end of the chain, so the system is in the Kitaev-like topological phase.

Next we check the $V\neq 0$ case. Using similar techniques, the topological number $N_2$ for this sublattice asymmetric case is (see the Appendix)
\begin{equation}\label{}
  N_2 = - \int_{-\pi}^{\pi} \frac{dq}{2\pi i} \partial_q \ln Z(q),
\end{equation}
where
\begin{equation}\label{}
  \begin{split}
     Z(q) =& Det M_2(q) = (V \cos \varphi)^2 + (g-h)(g^* + h^*) \\
       =& (V \cos \varphi)^2 + 2[ t^2(1+\lambda^2 \cos^2 \varphi) - \Delta^2 ] \\
       & + 2[ t^2(1-\lambda^2 \cos^2 \varphi) + \Delta^2 ]\cos q - 4i t\Delta \sin q.
  \end{split}
\end{equation}
Here, $N_2$ is the winding number of $Z(q)$, and is determined by the cross points of the real axis at $q=0$ and $\pi$. When $\Delta > 0$, we have
\begin{equation}\label{}
\begin{split}
     Z(0)Z(\pi) < 0 &\Rightarrow N_2=1, \\
     Z(0)Z(\pi) > 0 &\Rightarrow N_2=0,
\end{split}
\end{equation}
with
\begin{equation}\label{}
  \begin{split}
     Z(0) =& (V \cos \varphi)^2 + 4 t^2, \\
     Z(\pi) =& (V\cos \varphi)^2 + 4(t^2 \lambda^2 \cos^2 \varphi - \Delta^2).
  \end{split}
\end{equation}
When $\Delta<0$, we have $N_2=-1$ in the topological regime but it is equivalent to the $N_2=1$ phase. In Figs. \ref{fig5}(b) - Fig. \ref{fig5}(d), we present the phase diagrams of the system according to the topological number $N_2$. There are two phases when $V \neq 0$: $N_2 = \pm1$ is the Kitaev-like topological phase, and $N_2=0$ is the trivial phase. The phase boundary between these two phases is $(V\cos \varphi)^2 = 4[\Delta^2 - t^2 (\lambda \cos \varphi)^2]$, which is the gap closing condition we got earlier [see Eq. (\ref{eigenvalues})].

So, by checking the energy spectra and the topological numbers of the commensurate AAH model, we can completely describe the phases of the system. In addition, we can also tune the system into different topological phases by changing the system parameters such as the superconducting pairing amplitude and the modulation of the hopping amplitude, etc.

\section{Summary}\label{sec5}
In this paper, we have investigated a generalized Aubry-Andr\' e-Harper model with superconducting pairing both in the incommensurate and commensurate case. Due to the modulations in the hopping amplitude and the on-site potential, this model shows aboundant physical phenomena as we tune the system parameters.  When the lattice is incommensurate [$\alpha=(\sqrt{5}-1)/2$], the system shows a phase transition from extended states to localized states when the disordered potential becomes larger than some critical value. This critical value as well as the critical region will be reduced due to the modulation in the hopping amplitude especially when this modulation become strong. On the other hand, in the commensurate lattice case ($\alpha=1/2$), the system presents three different phases in different parameter regimes: a SSH-like trivial phase, a SSH-like topological phase, and a Kitaev-like topological phase. The topological numbers and the phase diagram are illustrated in this work. This generalized AAH model combined with superconducting pairing unifies those different phases and can provide an ideal test field for different kinds of phenomena. All these results can be generalized to discuss other cases with different $\alpha$ values, both rational and irrational.

As proposed in Ref. \cite{Ganeshan} (see the Supplemental Material there), a generalized AAH model with modulations in the hopping amplitude could be realized in cold atom systems. The generalized Aubry-Andr\'e-Harper model we propose here, though more complicated, might also be realized in cold atom optical lattices \cite{Billy, Roati}, or semiconductor structures \cite{Sarma, Merlin} in the future. This generalized AAH model may also have potential applications in understanding the topological properties of other similar systems, as discussed in Refs. \cite{Li, Ganeshan2}.

\section*{Acknowledgments}
This work has been supported by the NSFC under Grant No. 11274195 and the National Basic Research Program of China (973 Program) Grants No. 2011CB606405 and No. 2013CB922000. S. C. is supported by NSFC under Grants No. 11425419, No. 11374354 and No. 11174360, and the Strategic Priority Research Program (B) of the Chinese Academy of Sciences (No. XDB07020000).

\section*{Appendix}
\setcounter{equation}{0}
\renewcommand{\theequation}{{A}.\arabic{equation}}
In this Appendix, we investigate the topological class and calculate the topological numbers of the system. This part is mainly following the discussions in Ref. \cite{Wakatsuki}. Our model is time-reversal symmetric, $T \mathcal{H} (q) T^{-1} =  \mathcal{H} (-k)$,  since the coefficients $V$, $t$, $\lambda$ and $\Delta$ are all real. The time-reversal operator is defined as $T=K$, which takes the complex conjugate. When $V=0$, the system has sublattice symmetry. The operator for sublattice symmetry is defined by
\begin{equation}\label{}
  C_1 = \sigma_z =
  \begin{pmatrix}
    1 & 0 & 0 & 0 \\
    0 & -1 & 0 & 0 \\
    0 & 0 & 1 & 0 \\
    0 & 0 & 0 & -1
  \end{pmatrix},
\end{equation}
where $\sigma_z$ is the Pauli matrix acting on the sublattice degree of freedom. It can be shown that $C_1 \mathcal{H} (q) C_1^{-1} = -\mathcal{H}(q)$. Since we have $T^2=1$ and $C_1^2 = 1$, the topological class of the system is BDI.

If $V \neq 0$, the system will no longer have sublattice symmetry anymore. However, due to the superconducting pairing, particle-hole symmetry can still be guaranteed. The particle-hole operator is defined by $P=\tau_x K$, with $\tau_x$ being the Pauli matrix acting on the particle-hole space. The Hamiltonian satisfies $P \mathcal{H} (q) P^{-1} = -\mathcal{H} (-q)$. We can introduce the chiral operator as
\begin{equation}\label{}
  C_2 = TP = \tau_x =
  \begin{pmatrix}
    0 & 0 & 1 & 0 \\
    0 & 0 & 0 & 1 \\
    1 & 0 & 0 & 0 \\
    0 & 1 & 0 & 0
  \end{pmatrix}.
\end{equation}
Then we have $C_2 \mathcal{H}(q) C_2^{-1} = -\mathcal{H} (q)$ and $C_2^2 = 1$, so the topological class is still BDI. The 1D system in the BDI class is characterized by the $\mathbb{Z}$ index.

Now we calculate the topological numbers of the system. Firstly, we consider the sublattice symmetric case ($V=0$) whose topological number is defined as
\begin{equation}\label{}
  N_1 = Tr \int_{-\pi}^{\pi} \frac{dq}{4\pi i} C_1 g^{-1} \partial_q g,
\end{equation}
where $g(q)=-\mathcal{H}^{-1} (q)$ is the Green's function at zero energy. $N_1$ is equivalent to the chiral index. After introducing a unitary transformation
\begin{equation}\label{}
  U_1 =
  \begin{pmatrix}
    1 & 0 & 0 & 0 \\
    0 & 0 & 1 & 0 \\
    0 & 1 & 0 & 0 \\
    0 & 0 & 0 & 1
  \end{pmatrix},
\end{equation}
we have
\begin{equation}\label{}
  U_1 C_1 U_1^\dagger = \tau_z, \hspace{1cm} U_1 \mathcal{H} U_1^\dagger =
  \begin{pmatrix}
    0 & M_1 \\
    M_1^\dagger & 0
  \end{pmatrix},
\end{equation}
with
\begin{equation}\label{}
M_1=
  \begin{pmatrix}
    g & h \\
    -h & -g
  \end{pmatrix}.
\end{equation}
where $g(q)$ and $h(q)$ are defined earlier. Then the chiral index is given by
\begin{equation}\label{}
  \begin{split}
     N_1= & -Tr \int_{-\pi}^{\pi} \frac{dq}{2\pi i} M_1^{-1} \partial_q M_1 \\
       =& - \int_{-\pi}^{\pi} \frac{dq}{2\pi i} \partial_q \ln Det[M_1]\\
       =& - \int_{-\pi}^{\pi} \frac{dq}{2\pi i} \partial_q \ln [(h-g)(h+g)]
  \end{split}
\end{equation}
with
\begin{equation}\label{}
  \begin{split}
      h -g =& [t (1- \lambda \cos \varphi) - \Delta] + [ t(1+\lambda \cos \varphi) + \Delta ]e^{-iq}, \\
     h+g =& [-t(1-\lambda \cos \varphi) - \Delta] - [t(1+\lambda \cos \varphi) - \Delta] e^{-iq}.
  \end{split}
\end{equation}
So the topological number $N_1$ is
\begin{equation}\label{}
  N_1 = \Theta ( t \lambda \cos \varphi - \Delta) + \Theta (t\lambda \cos \varphi + \Delta )
\end{equation}
with $\Theta(x)$ being the step function.

Next we check the $V\neq 0$ case. The topological number $N_2$ associated with the chiral operator $C_2$ is
\begin{equation}\label{}
  N_2 = Tr \int_{-\pi}^{\pi} \frac{dq}{4\pi i} C_2 g^{-1} \partial_q g.
\end{equation}
Considering a unitary transformation
\begin{equation}\label{}
  U_2= \frac{1}{\sqrt{2}}
  \begin{pmatrix}
    1 & 0 & 1 & 0 \\
    0 & 1 & 0 & 1 \\
    -i & 0 & i & 0 \\
    0 & -i & 0 & i
  \end{pmatrix},
\end{equation}
which corresponds to the Majorana fermion representation
\begin{equation}\label{}
  c_j = \frac{1}{2}(\gamma_{j,1} + i \gamma_{j,2}), \hspace{1cm} c_j^\dagger = \frac{1}{2}(\gamma_{j,1} - i \gamma_{j,2}).
\end{equation}
It is easy to check that
\begin{equation}\label{}
  U_2 C_2 U_2^\dagger = \tau_z, \hspace{1cm} U_2 \mathcal{H}U_2^\dagger =
  \begin{pmatrix}
  0 & M_2 \\
  M_2^\dagger & 0
  \end{pmatrix},
\end{equation}
with
\begin{equation}\label{}
M_2=
  \begin{pmatrix}
    -i V\cos \varphi & i(g-h) \\
    i(g^* + h^*) & i V\cos \varphi
  \end{pmatrix}.
\end{equation}
Thus we have
\begin{equation}\label{}
  N_2 = -Tr \int_{-\pi}^{\pi} \frac{dq}{2\pi i} M_2^{-1} \partial_q M_2 = - \int_{-\pi}^{\pi} \frac{dq}{2\pi i} \partial_q \ln Z(q),
\end{equation}
where
\begin{equation}\label{}
  \begin{split}
     Z(q) =& Det M_2(q) = (V \cos \varphi)^2 + (g-h)(g^* + h^*) \\
       =& (V \cos \varphi)^2 + 2[ t^2(1+\lambda^2 \cos^2 \varphi) - \Delta^2 ] \\
       & + 2[ t^2(1-\lambda^2 \cos^2 \varphi) + \Delta^2 ]\cos q - 4i t\Delta \sin q.
  \end{split}
\end{equation}
$N_2$ is the winding number of $Z(q)$, and is determined by the cross points of the real axis at $q=0$ and $\pi$. For $\Delta > 0$, we find
\begin{equation}\label{}
\begin{split}
     Z(0)Z(\pi) < 0 &\Rightarrow N_2=1, \\
     Z(0)Z(\pi) > 0 &\Rightarrow N_2=0,
\end{split}
\end{equation}
with
\begin{equation}\label{}
  \begin{split}
     Z(0) =& (V \cos \varphi)^2 + 4 t^2, \\
     Z(\pi) =& (V\cos \varphi)^2 + 4(t^2 \lambda^2 \cos^2 \varphi - \Delta^2).
  \end{split}
\end{equation}
For $\Delta<0$, we have $N_2=-1$ in the topological regime but it is equivalent to the $N_2=1$ phase.

{}

\end{document}